\documentclass[%
 twocolumn,
 amsmath,amssymb,
 aps
]{revtex4}

\usepackage{gensymb}
\usepackage{graphicx}
\usepackage{float}
\usepackage{dcolumn}
\usepackage{bm}
\usepackage{epstopdf}
\usepackage{array}
\usepackage{bigstrut}
\usepackage{longtable}
\usepackage{rotating,booktabs}
\usepackage{booktabs,threeparttable}
\usepackage{color}
\usepackage[colorlinks,linkcolor=blue, citecolor=green, urlcolor=blue, anchorcolor=blue]{hyperref}
\begin{document}


\title{Measurement of hyperfine structure and the Zemach radius in $\rm^6Li^+$ using optical Ramsey technique}

\author{Wei Sun,$^{1,2,*}$ Pei-Pei Zhang,$^{1,*}$ Peng-peng Zhou,$^{1,2,5}$ Shao-long Chen,$^{1,2}$ Zhi-qiang Zhou,$^{1,2,5}$ Yao Huang,$^{1,2}$  Xiao-Qiu Qi,$^{6}$ Zong-Chao Yan,$^{3,1}$ Ting-Yun Shi,$^{1}$ G. W. F. Drake,$^{4}$ Zhen-Xiang Zhong,$^{1}$ Hua Guan,$^{1,2,\dag}$ and Ke-lin Gao$^{1,2,\ddag}$}

\affiliation{%
\mbox{$^1$State Key Laboratory of Magnetic Resonance and Atomic and Molecular Physics,}\\
\mbox{Innovation Academy for Precision Measurement Science and Technology, Chinese Academy of Sciences, Wuhan 430071, China}\\
\mbox{$^2$Key Laboratory of Atomic Frequency Standards, Innovation Academy for Precision Measurement Science and Technology,}\\
\mbox{Chinese Academy of Sciences, Wuhan 430071, China}\\
\mbox{$^3$Department of Physics, University of New Brunswick, Fredericton, New Brunswick, Canada E3B 5A3}\\
\mbox{$^4$Department of Physics, University of Windsor, Windsor, Ontario, Canada N9B 3P4}\\
\mbox{$^5$University of Chinese Academy of Sciences, Beijing 100049, China}\\
\mbox{$^6$Key Laboratory of Optical Field Manipulation of Zhejiang Province and}\\
\mbox{Physics Department of Zhejiang Sci-Tech University, Hangzhou 310018, China}\\
}

\date{\today}

\begin{abstract}
{We investigate the $2\,^3\!S_1$--$2\,^3\!P_J$ ($J = 0, 1, 2$) transitions in $\rm^6Li^+$ using the optical Ramsey technique and achieve the most precise values of the hyperfine splittings of the $2\,^3\!S_1$ and $2\,^3\!P_J$ states, with smallest uncertainty of about 10~kHz. The present results reduce the uncertainties of previous experiments by a factor of 5 for the $2\,^3\!S_1$ state and a factor of 50 for the $2\,^3\!P_J$ states, and are in better agreement with theoretical values. Combining our measured hyperfine intervals of the $2\,^3\!S_1$ state with the latest quantum electrodynamic (QED) calculations, the improved Zemach radius of the $^6$Li nucleus is determined to be 2.44(2)~fm, with the uncertainty entirely due to the uncalculated QED effects of order $m\alpha^7$. The result is in sharp disagreement with the value 3.71(16) fm determined from simple models of the nuclear charge and magnetization distribution. We call for a more definitive nuclear physics value of the $\rm^6Li$ Zemach radius.}

\end{abstract}

\maketitle

\noindent

\textbf{Introduction.}
High precision spectroscopy of heliumlike ions such as Li$^+$ provides an important platform in the search for new physics beyond the standard
model~\cite{Pachucki2017,Drake2021}, and a unique measuring tool for nuclear properties~\cite{RMP2013}. The Zemach radius of the $^6$Li$^+$ isotope with spin $1$ is a particular case in point where there is a marked disagreement between the value obtained from nuclear structure models~\cite{Yerokhin2008}, and that derived from the atomic hyperfine structure (hfs) coupled with high precision atomic theory~\cite{Puchalski2013,Qi2020}.  The purpose of this Letter is to report the results of measurements of hfs that improve the experimental accuracy of previous measurements~\cite{Guan2020} greatly, and thereby sharpen the disagreement with the apparent value obtained from nuclear structure models~\cite{Yerokhin2008}.  The disagreement is puzzling since there is good agreement for the case of the second stable isotope $^7$Li$^+$ with spin $3/2$~\cite{Qi2020}.

\begin{figure}[tb]
\includegraphics[width=0.45\textwidth,trim=200 70 50 50,clip]{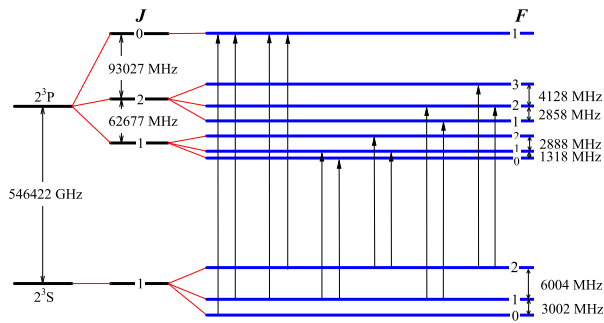}
\caption{\label{fig:energy_level_scheme} Fine and hyperfine structures for the $2\,^3\!S$ and $2\,^3\!P$ states of $\rm^6Li^+$ (not to scale). The six pairs of transitions marked by the arrows are the measured ones. 
}
\end{figure}

Hyperfine structure refers to the additional splitting of atomic energy levels due to nuclear spin. It is proportional to the magnetic $g$-factors of the electron and nucleus. The Zemach radius enters as a correction to the dominant Fermi contact term $E_{\rm F}$~\cite{Qi2020}.
The hfs can thus be expressed in the form~\cite{Riis1994,Puchalski2013}
\begin{equation}
H_{\rm hfs} = E_{\rm F}(1+H_{\rm HO}),
\end{equation}
where $H_{\rm HO}$ represents higher-order corrections given by
\begin{equation}
\label{deltaHO}
H_{\rm HO} = a_e+\delta_{\rm QED} - 2ZR_{\rm em}/a_0,
\end{equation}
with $a_e$ being the anomalous magnetic moment of the electron, $\delta_{\rm QED}$ a sum of higher-order QED corrections, $Z$ the nuclear charge, $a_0$ the Bohr radius, and $R_{\rm em}$
the nuclear Zemach radius defined in terms of the nuclear charge and magnetization densities
$\rho_{\rm e}({\bf r})$ and $\rho_{\rm m}({\bf r})$ by \cite{Zemach56}
\begin{equation}
R_{\rm em} = \int\int d^3r\,d^3r'\,\rho_{\rm e}({\bf r})\rho_{\rm m}({\bf r}')|{\bf r} - {\bf r}'|\ .
\end{equation}
In our previous work~\cite{Qi2020}, two values of $R_{\rm em}$ were obtained from two hfs intervals in $2\,^3\!S_1$, which was 2.40(16)~fm from $F=0 - F=1$ interval and 2.47(8)~fm from $F=1 - F=2$ interval,  see Fig.~\ref{fig:energy_level_scheme} for the energy level scheme of $\rm^6Li^+$. These two values are consistent with each other, but in disagreement with the value 3.71(16)~fm derived from simple nuclear models~\cite{Yerokhin2008}. The present work confirms the previous discrepancy and reduces the uncertainty in $R_{\rm em}$ by a factor of 4, with the dominant source of uncertainty being the $\delta_{\rm QED}$ term. The hfs measurements themselves are more accurate by a factor of 5 and 50 for $2\,^3\!S_1$ and $2\,^3\!P_J$, respectively. For the $2\,^3\!P_J$ states, no experimental hfs results with comparable precision are available. The latest hfs splitting measurement of $\rm^6Li^+$~\cite{Clarke2003} claims that the uncertainty is only a few hundred kHz, but is not in good agreement with theory.

The present measurements of hfs 
are based on the optical Ramsey technique~\cite{Baklanov1976,Borde1984}, which can be taken as an optical analog of the separated oscillatory field (SOF) or Ramsey method~\cite{Ramsey1950}  used in radiofrequency spectroscopy to reduce the transit-time broadening and thus the statistical uncertainty. Our experimental scheme is similar to that of Bergquist \emph{et al.}~\cite{Bergquist1977}. Briefly, a collimated $\rm^6Li^+$ ion beam is sent transversely through three consecutive equally-spaced standing-wave light fields, and the resulting emission line shape exhibits an interference pattern (Ramsey fringes), of which the period is determined by the travel time between adjacent radiation zones. By adjusting the spacing of the standing-wave light beams, Ramsey fringes with linewidth of about 5~MHz are obtained, which is close to the 3.7~MHz natural linewidth and narrower than those obtained previously~\cite{Bayer1979,Riis1994,Clarke2003} by at least one order of magnitude.

\begin{figure}[t]
\includegraphics[width=0.405\textwidth,trim=190 16 60 13,clip]{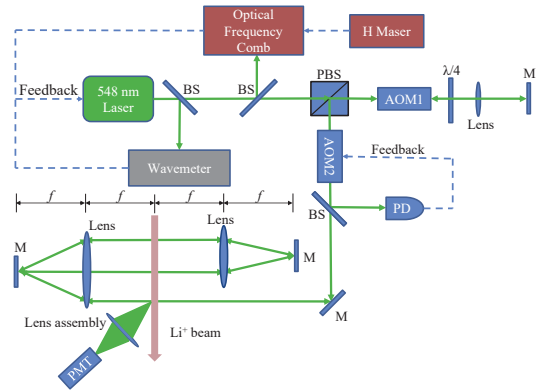}
\caption{\label{fig:experimental_setup} Schematic of the experimental setup. BS: beam splitter; PBS: polarized beam splitter; $\lambda$/4: quarter-wave plate; M: high reflecting mirror; AOM: acousto-optic modulator; PD: photodiode; PMT: photomultiplier tube; \textit{f}: lens focal length.}
\end{figure}

\noindent
\textbf{Experiment}. A schematic of the experiment is shown in Fig.~\ref{fig:experimental_setup}. A continuous beam of $2\,^3\!S_1$ metastable $\rm^6Li^+$ ions is produced by an ion beam source~\cite{Chen2019}. Briefly, heating a bulk lithium sample (95$\%$ $^6$Li, Sigma-Aldrich) in an oven produces an effusion beam of neutral lithium atoms, which intersects a focused electron beam. By impact, part of the lithium atoms are ionized. The $\rm^6Li^+$ ions, about 1.8$\%$ of which are in the $2\,^3\!S_1$ state, are not further separated. They are extracted from the impact region and collimated to a mono-velocity beam ($v\approx 1.3\times10^5$~m/s and $\delta v/v \sim 10^{-3}$) using an electrostatic lens system. The ion beam  with a diameter of about 5~mm, a beam current of 20~mA, and a beam divergence angle of 0.5~mrad enters into another vacuum chamber and interacts with the laser.

The 548~nm laser, used to excite the $2\,^3\!S_1$--$2\,^3\!P_J$ transition of Li$^+$, is generated by the second-harmonic generation of a 1097~nm narrow-band fiber laser (Y10, NKT Photonics) followed by optical amplification with a fiber amplifier. The laser has a linewidth of 10.8 kHz in free-running mode. It is frequency-stabilized by locking to a wavemeter (WS7-60, HighFinesse) and an optical frequency comb (FC8004, Menlo Systems)~\cite{Zhou2021}. The optical frequency comb with the repetition and carrier offset frequencies referenced to a GPS-disciplined hydrogen maser (CHI-75A, Kvarz) is also adopted to measure the laser's absolute frequency. A double-pass~\cite{Donley2005} acousto-optic modulator (AOM1 in FIG.~\ref{fig:experimental_setup}, MT200-AO-5-VIS, AA Opto Electronic), driven by a field programmable gate array (FPGA) referenced to the same hydrogen maser, is used to scan the laser frequency in a random order when a spectrum is being recorded. A second similar acousto-optic modulator is used to stabilize the intensity of the laser beam~\cite{Kim2007}.

Before entering the vacuum chamber, the frequency- and intensity-stabilized laser is collimated and guided through a half wave plate and a Glan-Taylor polarizer (DGL10, Thorlabs) with an extinction ratio greater than 100000:1. By rotating the wave plate together with the polarizer, the linear laser polarization angle can be adjusted. To create the required three equidistant standing-wave light fields, two cat’s-eye retroreflectors are employed, with a configuration shown in FIG.~\ref{fig:experimental_setup}. Each cat’s-eye consists of an antireflective achromatic lens of focal length $f = 500$~mm and a high reflecting mirror (R\textgreater 99.9$\%$) located at its focal point. The laser beam diameter is 2.3~mm and the spacing between adjacent beams is around 6~mm. The Laser beams enter the vacuum chamber via two optically flat fused-silica viewports with antireflective coating on both sides (R\textless 0.1$\%$). The background pressure of the chamber is $2\times 10^{-6}$~Pa, and is maintained with a turbomolecular pump (HiPace 300, Pfeiffer Vacuum). A lens assembly and a photomultiplier tube (PMT2100, Thorlabs) mounted in the direction orthogonal to the ion and laser beams are used to collect the fluorescence emitted by the $\rm^6Li^+$ ions after they pass transversely through the three laser radiation zones. The collection efficiency is roughly 10$\%$.

\begin{figure*}[tb]
\includegraphics[width=\textwidth,trim=0 350 0 0,clip]{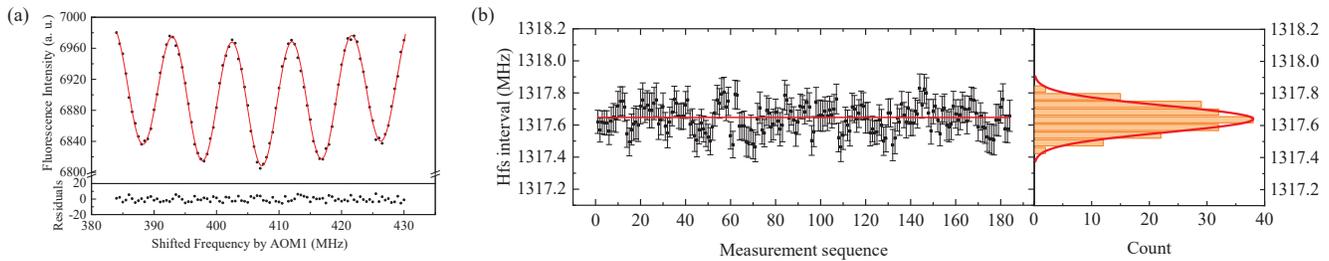}
\caption{\label{fig:spectrum} Measurement of the $2\,^3\!P_1^{0-1}$ interval in $\rm^6Li^+$. (a) Ramsey spectrum from a single scan of one of the measured transitions. The solid red line is a data fit to an exponentially damped sinusoidal function. Residuals of the fit are shown in the lower panel. (b) Experimental results for the $2\,^3\!P_1^{0-1}$ interval of $\rm^6Li^+$.}
\end{figure*}

Figure~\ref{fig:spectrum}(a) shows a sample optical Ramsey spectrum of the $2\,^3\!S_1$ ($F=1$) $\rightarrow$ $2\,^3\!P_1$ ($F=0$) transition recorded with the shifted laser frequency as the abscissa and the detected fluorescence intensity as the ordinate. 

For brevity, we denote the hfs interval between $F$ and $F'$ sublevels of a $2\,^{2S+1}\!L_J$ state as $2\,^{2S+1}\!L_J^{F-F'}$. As shown in Fig.~\ref{fig:energy_level_scheme}, every hfs interval can be determined by taking the difference of two neighboring transition frequencies in the Ramsey spectrum, measured 120 to 200 times and fitted to a line profile to determine the line center. Figure~\ref{fig:spectrum}(b) shows the measurements of the $2\,^3\!P_1^{0-1}$ interval of $\rm^6Li^+$, together with the histogram for 185 measurements. The results follow a normal distribution with a statistical uncertainty of 4~kHz. Results for the other hfs splittings are shown in Table~\ref{table:6Liexperimentalinterval} with uncertainties less than 10~kHz.

\noindent
\textbf{Results and discussion.} We have evaluated various systematic effects and the results are listed in Table~\ref{table:6Liexperimentalinterval}.

\begin{figure}[tb]
\includegraphics[width=0.5\textwidth,trim=70 240 100 90,clip]{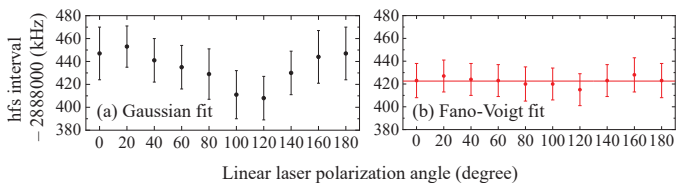}

\caption{\label{fig:quantum_interference} Dependence of the measured $2\,^3\!P_1^{1-2}$ interval of $\rm^6Li^+$ on laser polarization angle relative to the direction of the photodetector. (a) Points are obtained by fitting the envelope with a Gaussian function. (b) Points are obtained by fitting the envelope with a Fano-Voigt function.}
\end{figure}

\textit{Quantum interference.} Multiplets of atomic resonance lines are subject to mutual line pulling due to quantum interference~\cite{Low1952,Horbatsch2010,Udem2019}, which may introduce a systematic error in high precision spectroscopy~\cite{Horbatsch2011,Marsman2012,Hessels2012,Marsman2014,Marsman2015,Feng2015,Zheng2017,Brown2013,Beyer2017}. 
We have adopted an approximate model based on a Fano-Voigt line shape to reduce the uncertainty, as first developed by Udem \emph{et al.}~\cite{Udem2019}, and described in the Supplemental Material.
To test this model, we rotated the linear laser polarization and inspected the line pulling without/with (Fig.~\ref{fig:quantum_interference} (a/b)) taking into account the quantum interference. The results become nearly independent of laser polarization, showing that our approximate line shape is realistic.  The residual fluctuation of 8~kHz in Fig.~\ref{fig:quantum_interference}(b) was taken as the uncertainty.

\begin{figure}[tb]
\includegraphics[width=0.45\textwidth,trim=60 80 70 30,clip]{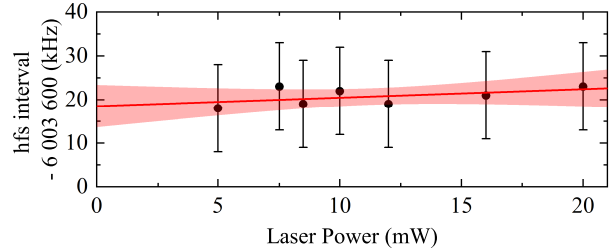}
\caption{\label{fig:frequencyvspower} Dependence of the measured $2\,^3\!S_1^{1-2}$ interval of $\rm^6Li^+$ on laser power. A linear fit is shown with the shaded area representing a 95$\%$ confidence band.}
\end{figure}

\textit{Power dependence.} The axial velocity distribution of the ions may cause a power-dependent shift in Ramsey fringes~\cite{Barger1981,Bava1981}. To assess this effect, measurements of the $2\,^3\!S_1^{1-2}$ hfs were carried out with laser power ranging from 5 to 20~mW, as shown in Fig.~\ref{fig:frequencyvspower}. The measured values did not show an obvious power dependence. A final result was obtained by extrapolation to the zero-laser-power limit and an uncertainty of 5~kHz was assigned. Other hfs intervals were all measured at laser power of 10~mW, without extrapolation-based correction, but were also assigned an uncertainty of 5~kHz.

\textit{First-order Doppler effect.} The optical alignment of the two cat’s-eye retroreflectors was performed by an interferometry method, leading to a parallelism of the counter-propagating waves within $1\times10^{-5}$~rad. This parallelism would introduce a first-order Doppler shift well below
1~kHz for a hfs interval. However, temperature fluctuation of 0.1~K during one measuring cycle might lead to a displacement of the cat’s-eye foci of about 0.7~$\mu$m, which in turn caused an angular variation of a retroreflected light by 0.03~$\mu$rad~\cite{Snyder1975}. A shift of 3.7~kHz was assigned as the uncertainty due to angular variation.

\textit{Second-order Doppler effect.} For our transverse excitation geometry, the second-order Doppler shift amounts to $\nu_0v^2/(2c^2)$ for ions of velocity $v$. With $v$ = $1.3\times10^5$~m/s and $\delta v/v = 10^{-3}$, the second-order Doppler shifts were calculated together with their estimated uncertainties, as listed in Table~\ref{table:6Liexperimentalinterval}.

\textit{Zeeman effect.} Under our residual geomagnetic field of $\sim$0.3~G, Zeeman components are not resolvable, leading to an extra line-broadening. We used linearly polarized light with purity better than 99$\%$. For the three-standing-wave configuration, a circularly polarized light traveling in one direction was always balanced by an oppositely polarized light traveling in the opposite direction. Therefore, the Zeeman components always appeared in the spectrum as symmetric pairs and did not shift the center of the line as long as the optical loss in the beam path was negligible. Taking the optical loss as 1$\%$, the shift caused by the imperfect balance of Zeeman component intensity was calculated and assigned as the uncertainty due to the Zeeman effect. The measured hfs values with circularly polarized light were found to be consistent. Other systematic effects due to instabilities of the optical frequency comb, photon recoil, collisional effects, and patch charges were considered and found to be negligible.

\begin{widetext}
\begin{center}
\begin{table*}[tb]
\caption{\label{table:6Liexperimentalinterval}Experimental results for the hfs intervals of the $2\,^3\!S_1$ and $2\,^3\!P_J$ states of $\rm^6Li^+$ with uncertainty budget, in kHz.}
\small
\begin{ruledtabular}
\begin{tabular}{lrrrrrr}
\textrm{Source of error}&\textrm{$2\,^3\!S_1^{0-1}$}&\textrm{$2\,^3\!S_1^{1-2}$}&\textrm{$2\,^3\!P_1^{0-1}$}&\textrm{$2\,^3\!P_1^{1-2}$}&\textrm{$2\,^3\!P_2^{1-2}$}&\textrm{$2\,^3\!P_2^{2-3}$}\\
\colrule
Statistical & 3 001 783 (6) & 6 003 618 (4) & 1 317 652 (6) & 2 888 423 (4) & 2 858 019 (6) & 4 127 891 (4)\\
Quantum interference & (8) & (8) & (8) & (8) & (8) & (8)\\
Laser power & (5) & (5) & (5) & (5) & (5) & (5)\\
1$^{\rm st}$-order Doppler effect & (3.7) & (3.7) & (3.7) & (3.7) & (3.7) & (3.7)\\
2$^{\rm nd}$-order Doppler effect & 0.27(1) & 0.54(3) & 0.12(1) & 0.26(1) & 0.26(1) & 0.37(2)\\
Zeeman effect & (0.6) & (0.3) & (0.5) & (0.3) & (0.4) & (0.2)\\
Total & 3 001 783 (12) & 6 003 619 (11) & 1 317 652 (12) & 2 888 423 (11) & 2 858 019 (12) & 4 127 891 (11)\\
\end{tabular}
\end{ruledtabular}
\end{table*}
\end{center}

\begin{center}
\begin{table*}[tb]
    \scriptsize\caption{\label{table:comparison} Experimental and theoretical hfs intervals of the $2\,^3\!S_1$ and $2\,^3\!P_J$ states of $\rm^6Li^+$, in MHz. For the theoretical calculation in this work, the nuclear electric quadrupole moment used is $-0.0806(6)$ fm$^2$~\cite{Stone2016} and the Zemach radius is 2.44(2) fm which is derived in this work.}
     \small
     \begin{ruledtabular}
     \begin{tabular}{ccccccc}
     \multicolumn{1}{c}{}
    &\multicolumn{3}{c}{Experiment}
    &\multicolumn{3}{c}{Theory}\\
    \cline{2-4} \cline{5-7}
    $ $ & $ $ & $ $ & $ $ & Drake and  & $ $ & $ $\\
     hfs intervel & Kowalski {\it et al.}~\cite{Kowalski1983} & Clarke {\it et al.}~\cite{Clarke2003} & This work & co-workers~\cite{Riis1994} & Qi {\it et al.}~\cite{Qi2020} & This work\\
    \hline
     $2\,^3\!S_1^{0-1}$ & 3001.780(50) & 3001.83(47) & 3001.783(12) & 3001.765(38) & $ $ & $ $\\
     $2\,^3\!S_1^{1-2}$ & 6003.600(50) & 6003.66(51) &  6003.619(11) & 6003.614(24) & $ $ & $ $\\
     $2\,^3\!P_1^{0-1}$ & $ $ & 1316.06(59) & 1317.652(12) & 1317.649(46) & 1317.732(31) & 1317.736(15)\\
     $2\,^3\!P_1^{1-2}$ & $ $ & 2888.98(63) & 2888.423(11) & 2888.327(29) & 2888.379(20) & 2888.391(10)\\
     $2\,^3\!P_2^{1-2}$ & $ $ & 2857.00(72) & 2858.019(12) & 2858.002(60) & 2857.962(43) & 2857.972(21)\\
     $2\,^3\!P_2^{2-3}$ & $ $ & 4127.16(76) & 4127.891(11) & 4127.882(43) & 4127.924(31) & 4127.937(15)\\
    \end{tabular}
    \end{ruledtabular}
\end{table*}
\end{center}
\end{widetext}

We present the experimental results for the hfs intervals
of $^6$Li$^+$ in Table~\ref{table:6Liexperimentalinterval}, along with the uncertainty budget. These results represent the most precise experimental values to date, with uncertainties of about 10 kHz for all hfs intervals.

\begin{figure}[tbp]
\includegraphics[width=0.5\textwidth]{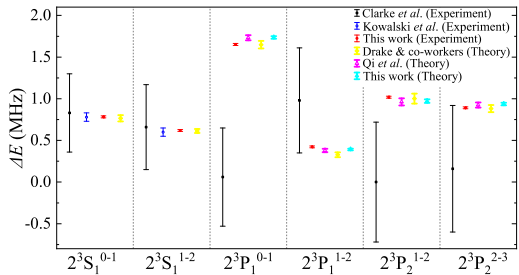}
\caption{\label{fig:intervalcomparison} Comparison of hfs splittings in the $2\,^3\!S_1$ and $2\,^3\!P_J$ states of $\rm^6Li^+$. $\Delta E$ stands for the six indicated intervals with offset of 3001, 6003, 1316, 2888, 2857, and 4127 MHz, respectively. The black, blue, and red lines represent the experimental results of Clarke \emph{et al.}~\cite{Clarke2003},  Kowalski \emph{et al.}~\cite{Kowalski1983}, and this work, respectively. The yellow and magenta lines represent the calculations of Drake and co-workers~\cite{Riis1994} as well as Qi \emph{et al.}~\cite{Qi2020}, respectively. The cyan lines represent calculations using the same theory as Ref.~\cite{Qi2020} and the Zemach radius of 2.44(2)~fm derived in this work.}
\end{figure}

Table~\ref{table:comparison} and Fig.~\ref{fig:intervalcomparison} show a comparison of our measured hfs intervals with other experimental and theoretical ones. For the $2\,^3\!S_1$ state, our experimental results not only are in good accord with those of Kowalski {\it et al.} using laser-microwave spectroscopy~\cite{Kowalski1983} and those of Clarke {\it et al.} using an electro-optic modulation technique~\cite{Clarke2003}, but also have much smaller uncertainties. In Ref.~\cite{Qi2020}, we derived the Zemach radius of the $^6$Li nucleus from the measured hfs splittings in the $2\,^3\!S_1$ state from Ref.~\cite{Kowalski1983}. The $2\,^3\!S_1^{0-1}$  and $2\,^3\!S_1^{1-2}$ intervals yielded 2.40(15)(6)~fm and 2.47(7)(2)~fm respectively for the Zemach radius, where the first uncertainty was from experiment and the second from QED theory (the same hereinafter). The newly measured same splittings combined with the same theory now yield 2.40(4)(7) fm and 2.44(1)(2) fm respectively. The relative contributions to the uncertainties from $\delta_{\rm QED}$ and the Zemach radius term in Eq.\ (\ref{deltaHO}) are 413(5) ppm and $-277(5)$ ppm respectively.  Thus, we take 2.44(2)~fm as the recommended value for the Zemach radius of the $^6$Li nucleus, where the uncertainty comes entirely from the uncalculated QED term of order $m\alpha^7$. In comparison, our Zemach radius for the Li$^+$ ion is in perfect agreement with the value 2.44(6) fm \cite{Qi2020} from the hfs of neutral $^6$Li derived by Puchalski and Pachucki~\cite{Puchalski2013} using the most recent measurements of Li {\it et al.}~\cite{Li2020}.

Using 2.44(2)~fm as our recommended Zemach radius for the $^6$Li nucleus, we recalculated the hfs intervals in the $2\,^3\!P_J$ states and displayed the results in Table~\ref{table:comparison}. Comparing with the most recent measurement of Ref.~\cite{Clarke2003}, our measured results have significantly improved their values and are in better agreement with theory, as is clearly seen in Fig.~\ref{fig:intervalcomparison}. However, there are still some small discrepancies between theory and experiment, which may imply that either some systematic experimental errors have not been properly accounted for, or the contribution from higher-order corrections of order $m\alpha^7$ have been underestimated. We also checked whether the assumed value of $-0.0806(6)$~fm$^2$~\cite{Stone2016}  for the electric quadrupole moment $Q_d$ might have a significant effect. A fit with $Q_d$ as an adjustable parameter yielded a value of low accuracy ($-0.38(20)$ fm$^2$) by taking into consideration
the uncertainties in the measured hfs, the determined Zemach radius, and the high-order QED term.

\textbf{Conclusion.}
We have demonstrated that the optical Ramsey technique can drastically reduce the transit-time broadening in spectra of Li$^+$ beam. To the best of our knowledge, this is the first time the optical Ramsey technique has been applied to a charged particle beam. This technique may find wider applications in future high-precision spectroscopy of charged particle beams.

The resulting improved measurements of the hfs intervals in the $2\,^3\!S_1$ and $2\,^3\!P_J$ states of $\rm^6Li^+$ are now sensitive to QED effects of order $m\alpha^7$.
The derived Zemach radius of 2.44(2)~fm for the $^6$Li nucleus is in sharp disagreement with the value 3.71(16) fm obtained from models of the nuclear charge and magnetization distributions \cite{Yerokhin2008}, in contrast to the good agreement for the case of $^7$Li.  It seems clear that the nuclear structure of $^6$Li is anomalous and requires further study.

\section*{Acknowledgments}
The authors thank Krzysztof Pachucki for reading the manuscript and providing insightful advice.
The authors also thank Chaohong Lee, Jiahao Huang, Shuiming Hu, Yu Robert Sun, and Hongping Liu for helpful discussions. We also thank Qunfeng Chen, Yanqi Xu, and Huanyao Sun for technical support in optical frequency comb and electronic circuits. This work is supported jointly by the National Natural Science Foundation of China (Grant Nos.\ 11934014, 11804373, 11974382, 12121004, 12174400 and 92265206), the Strategic Priority Research Program of the Chinese Academy of Sciences (Grant No. XDB21010400), CAS Youth Innovation Promotion Association (Grant Nos.\ Y201963 and 2018364), CAS Project for Young Scientists in Basic Research (Grant No.\ YSBR-055) and K. C. Wong Education Foundation (Grant No.\ GJTD-2019-
15). ZCY and GWFD acknowledge research support by the Natural Sciences and Engineering Research Council of Canada.
\\
$^*$These authors contributed equally to this work.\\
$^{\dag}$Email Address: guanhua@wipm.ac.cn\\
$^{\ddag}$Email Address: klgao@wipm.ac.cn

\providecommand{\noopsort}[1]{}\providecommand{\singleletter}[1]{#1}%

\end{document}